\documentclass[preprint]{aastex}

\shorttitle{Young Massive Stellar Cluster}
\shortauthors{Roman-Lopes et al.}

\begin{document}


\title{Discovery of a Young Massive Stellar Cluster associated with IRAS source 16177-5018 
\footnote{Based on observations made at Laborat\'orio Nacional de Astrofisica/MCT, Brazil}}


\author{A. Roman-Lopes, Z. Abraham and J. R. D. L\'epine}
\affil{Instituto de Astronomia, Geof\'\i sica e Ci\^encias Atmosf\'ericas, Universidade de
S\~ao Paulo \\ Rua do Mat\~ao 1226, 05508-900, S\~ao Paulo, SP, Brazil}

\email{roman@astro.iag.usp.br}


\begin{abstract}
We report the discovery of a young massive stellar cluster embedded in an extended HII region, invisible
at optical wavelengths where the extinction is $A_V \approx 28$ magnitudes, associated with the IRAS source 16177-5018. 
$J, H$ and nb$K$ imaging photometry combined with the $K_S$ 2MASS data show the presence of sources with infrared excess
emission at 2.2 $\mu$m, concentrated in an area of about one square parsec around a massive young stellar object
identified as the IRAS source. 
This object has a near-mid infrared spectral index betweem 2.2 and 25 $\mu$m $\alpha({\rm IR}) = d\;{\rm log}(\lambda F_\lambda)/d\; {\rm log} \lambda $
=4.78, characteristic of compact H II regions, with bolometric luminosity, inferred from the 
integrated near to far-infrared flux density of $2.8 \times 10^5 L_\odot$, which
corresponds to a ZAMS star of about $42 M_\odot$. From the color-magnitude diagram we were able to classify the majority 
of the cluster members as reddened massive stars earlier than spectral type B5. 
\end{abstract}




\section{Introduction}

Young massive stars are formed in dense giant molecular clouds, where the gas density  may be as 
high as $n_H  = 10^5-10^6$ cm$^{-3}$ and the temperature as low as $T = 10-20$ K. For this reason, 
the star forming regions that eventually lie in the core of the molecular cloud can be completely obscured at 
optical wavelengths by many magnitudes of extinction. On the other hand, the  dense
molecular clouds can be traced by molecular emission, like CS and NH$_3$ lines at radio frequencies, which 
are powerful tools for the determination of the physical conditions in the clouds. 
While OB stars are embedded in a molecular cloud, they dissociate and ionize the gas, forming
compact HII regions seen at radio 
wavelengths, and heat the circumstellar dust which eventually radiates most of the stellar luminosity in 
the far-infrared (FIR). Therefore, the most prominent tracers of young massive stars are also compact, 
or ultracompact 
H II (UC H II) regions, which have electron densities greater than $10^4$ cm$^{-3}$ and diameters smaller than 
 0.1 pc. Wood \& Churchwell (1989) showed that these ultracompact H II regions have characteristic 
spectral energy distributions in the far-infrared, occupying a well delimited region in the 
${\rm log}[F_\nu(60\mu m)/F_\nu(12\mu m)] \times {\rm log}[F_\nu(25\mu m)/F_\nu(12\mu m)]$ space. Bronfman et al. (1996) 
made an extensive survey of the CS (2-1) 
line emission toward IRAS point sources with colors characteristics of ultra-compact H II regions  
and Roman-Lopes et al. (2002) detected ammonia emission in the direction of the strongest
sources  in that survey.
With the advent of the large bidimensional near-infrared array detectors, the morphological 
and photometric studies of extremely young galactic stellar clusters were greatly benefited. At 
near-infrared wavelengths (1 to 2.5 $\mu$m) it is possible to probe deep into the dense dust clouds where star 
formation is taking place. At those wavelengths, very young objects present large 
infrared excess due the presence of warm circunstellar dust. 

In this paper we present the discovery of a young stellar cluster of massive stars and its
associated HII region, in the direction of the IRAS source 16177-5018. We combined 2MASS (Two Micron All Sky Survey) data with 
new NIR observations from Laborat\'orio Nacional de Astrofisica (LNA), Brazil. This work is a part of a survey aimed 
to the identification of stellar populations in the direction of IRAS sources that have colors characteristics
of ultracompact HII regions \citep{wood89} and strong CS (2-1) line emission \citep{bronf96}. 
The studied region is part of the RCW 106 complex, 
located in the southern Galactic plane at a distance of 3.8 kpc \citep{cas87}; it presents compact far 
infrared sources \citep{kar01}, intense NH$_3$ (J,K) = (1,1) line emission \citep{rom02} and continuum radio emission at 
5 GHz as well as  hydrogen recombination lines \citep{cas87}.

\section{Observations and data reduction}

The imaging observations were performed in June 2001 with the Near Infrared Camera (CamIV) of 
Laborat\'orio Nacional de Astrofisica, Brazil,  equipped with a Hawaii 1024x1024 pixel 
HgCdTe array detector mounted on the 0.6 m Boller \& Chivens telescope. The plate scale was 0.47 
arcsec/pixel and the mean values of the PSF full width at half maximum (MFWHM) were 1.2, 1.5 and 2 
arcsec at the $\it{J}$, $\it{H}$ and $\it{nbK}$ images, respectively.
The observations consisted of  8'x 8' frames in the direction of the IRAS source I16177-5028 using 
$\it{J, H}$ and $\it{nbK}$ filters. The narrow band $\it{K}$ filter is centered at 2.24956 $\mu$m with a FWHM of  
0.06333 $\mu$m, covering only the continuum emission. The total integration time was 540 s for the $\it{J}$ and 
$\it{H}$ 
bands  and 1620 s for the $\it{ nbK}$ filter, resulting in a sensitivity at 3$\sigma$ of 17.5, 16.8 and 13.4 magnitudes, respectively. 
The frames were dithered by 60 arcsec,
in order to remove bad pixels, cosmic rays and to eliminate the presence of objects 
with extended emission on the construction of the sky images. Individual frames were dark 
subtracted and flat-fielded before being combined. We obtained a set of "dome flats" for each 
filter using an illuminated white spot in the telescope dome. 
Sky frames were generated from a median-filtered set of the flattened frames in each band. To produce 
a combined image from each dithered set of images, we aligned the frames using the IRAF\footnote{IRAF is distributed by the 
National Optical Astronomy Observatories, which is operated by the Assotiation of Universities for Research in 
Astronomy, Inc. under contract to the National Science Foundation} 
subroutines, added the aligned frames and trimmed the resulting image to remove bad pixels at 
the edges of the frames. We used DAOFIND to locate stars 4 $\sigma$ above the local background and 
added to the DAOFIND list all stellar objects missed by this routine but found by visual inspection 
of each image. Because of source confusion within the region, photometry was obtained using 
the point spread function fitting algorithm ALLSTAR in the DAOPHOT package \citep{stet87}. For the $\it{J}$ and $\it{H}$ images 
the adopted PSF fitting radius were 2.0 and 3.2 pixels respectively and the PSF radius was 11 pixels. For the nb$\it{K}$ image
the PSF fitting radius was 4.2 pixels and the PSF radius was 13 pixels. The local sky was evaluated in an annulus with an inner radius of 
8 pixels and a width of 20 pixels.
The values of the completeness limits for $\it{J}, \it{H}$ and $\it{ nbK}$ bands are 16, 15.5 and 12.2 magnitudes
respectively and  were derived from the point where the number of detected sources of magnitude $m$, $N(m)$ 
deviates from a straight line in the log({\it N}) versus $m$ diagram as can be seen in figure 1.
Photometric calibration was done by observing standards from the list of Elias (1982), at several air masses before and after each 
set of integrations.  Direct equivalence was assumed between our $\it{nbK}$ magnitudes and the standard $\it{K}$ values. 

Photometry from 2MASS All Sky Point Source Catalogue \footnote{http://www.ipac.caltech.edu/cgi-bin/gator/nph-dd}
in the $\it{J}, \it{H}$ and $\it{K_S}$ filters became available recently. The $K_{S}$ 
filter, centered at 2.17 $\mu$m, has a bandpass of 0.32 $\mu$m. 
The completeness limits for this survey are 15.8, 15.1 and 14.3 magnitudes at $\it{J}$,  $\it{H}$ and $\it{K}$$_{S}$,
 respectively \citep{egan01}. 
 Since the 2MASS $K_{S}$ band photometry has a completeness limit greater than our nbK photometry, 
we decided to use the 2MASS catalogue astrometry and magnitudes in this filter to study the stellar population. Even though the 2MASS
survey was conducted with a 1.3 m telescope, the sensitivity and resolution of our IR camera compensates our
smaller collecting area, resulting in more detection in the $J$ and $H$ bands, as it will be shown in section 3.2.
For this reason we only used the more sensitive 2MASS $K_S$ photometry.

We also checked the consistence of our photometry comparing it with the data of the 2MASS survey. We took all point sources 
detected by 2MASS in an area of about 50 square
arcmin around IRAS16177-5018 and constructed a list of sources common to both
surveys in the three filters. This procedure gave us a total of 612, 781 and 232  sources in the $\it{J, H}$ and 
$\it{K}$ bands respectively.
We constructed graphs for M$_{ 2mass}$ versus M$_{CamIV}$ $\it{J, H}$ and $\it{K}$ magnitudes respectively, which
 are shown in figure 2. We see a good linear relation between the two systems, with a slope of 1 and
a dispersion that increases with the magnitude.

\section{Results \& Discussion}

The infrared images, especially in the $\it{H}$, nb$\it{K}$ and $\it{K_S}$ bands, show the presence of a small infrared 
nebula 
($\sim 70^{\prime\prime} \times 70^{\prime\prime}$  square arcsec) at the IRAS coordinate, with an apparent 
concentration of embedded 
stars, suggesting the presence of a cluster.  
In figure 3 we show a combined pseudoreal 
colour ($\it{J}$ represented in blue, $\it{H}$ in green and nb$\it{K}$ in red) image of the whole field, and 
amplified images of the nebular region in the $\it{J, H}$ and nb$\it{K}$ bands. 

In figure 4 we present the ($J-H$) versus ($H-K_S$) diagram for all stars detected in  the LNA $J$ and $H$ images and
in  the 2MASS $K_S$ image
or only at $H$ and $K_S$, using the completeness limit for the $J$ magnitude, together with the position of the 
main sequence, 
giant branch and reddening vectors for early and late type stars. 
The majority of the sources have colors of reddened photospheres but some of them appear to the right 
of the reddening line for early type stars, showing "excess" at 2.2 $\mu$m. When we combine the 
results shown by the color-color diagram with the spatial distribution of the sources, we verify 
that all stars situated to the right of the early type reddening vector  with  $H-K_S \geq 1.0$ are located 
in the nebular region.

It is well established that very young pre-main sequence objects present large infrared excess due 
to the presence of warm circumstellar dust.
Our result suggests that the stellar population in the direction of the IRAS source 
is very young, and  from their position in the $(J-H)$ versus $(H-K_S)$ diagram they are probably young massive
stellar objects.  

In order to estimate the amount of interstellar extinction in the direction of the IRAS source we 
computed the average value of the color excess $E(H-K)$ using the 11 sources from Table 1
(IRS 1, 2, 7, 8, 11, 18, 22, 23, 26, 28 and 31) that lie along 
the reddening line for early type stars in the color-color diagram. 
We  assumed that  the intrinsic $(H-K)_{0}$  colors are those of  luminous early type stars (-0.05 
for O6-O8V types as given by Koornneef, 1983). The mean color excess obtained was ${\it <E(H-K)>} = 1.75$, which
for a standard reddening law \citep{riek85} corresponds to a median visual extinction $<A_{\it{V}}>\approx 28$. 

\subsection{The IRAS source}

In Figure 5 we present a contour map constructed from the LNA nb$K$ image, which shows the region around IRAS16177-5018. 
The IRAS coordinate  has an intrinsic error  delimited by the ellipse 
plotted in the figure. Any object inside the ellipse could be identified with the IRAS source and in fact, more 
than one of these objects may be contributing to the detected infrared flux .  
A more accurate position for the IR source was obtained from the
Midcourse Space Experiment - MSX source catalog \footnote{
http://www.ipac.caltech.edu/ipac/msx/msx.html}. The MSX surveyed the entire 
Galactic plane within 
$\mid b \mid \leq 5^\circ$ in four mid-infrared spectral bands centered at 8.28, 
12.13, 14.65 and 21.34 $\mu$m, with image resolution of 20 arcsec and a global absolute astrometric accuracy of about
1.9 arcsec \citep{price01}. 
We found only one MSX source, with coordinates $\alpha(\rm{J2000)=16^{h}21^{m}31.4^{s}}$, 
$\delta(\rm{J2000)=-50^{d}25^{m}04^{s}}$
within the IRAS uncertainty ellipse; it coincides with the star we
labeled IRS7 in Table 1.
In fact this is the only  MSX source in an area of about 9 square arcmin around the IRAS 
point source with  detected flux at the four bands. 

In Figure 6 we plotted the near to far-infrared spectral energy distribution of IRS7, 
without any correction for absorption.
We can see  that the MSX and IRAS flux densities agree
very well at the common wavelengths (12 and 21-25 $\mu$m), indicating that only one star
is responsible for the IRAS emission.
The spectral index in the near and mid-infrared (betweem 2.2-25 $\mu$m), defined as 
$\alpha=d \;{\rm log}(\lambda F_\lambda)/d\; {\rm log} \lambda$ 
is $\alpha=4.78$, characteristic of extremely young stellar objects in a earlier stage of evolution, 
when the star is deeply surrounded by a thick dusty envelope.

We integrated the observed flux density between 1.25 and 100 $\mu$m and, assuming a distance of 3.8 kpc,
we obtained a luminosity $L=2.8 \times 10^5 L_\odot$. Since most of the energy is emitted
in the infrared, this can be considered a lower limit to the bolometric luminosity of the
embedded star, reprocessed by a dusty envelope.
PMS models \citep{iben65} show that massive stars (above 9 M$_\odot$) evolve at constant luminosity from the
Hayashi track to the Main-Sequence. Therefore, the measured luminosity gives directly a lower limit
to the stellar mass of    42 M$_\odot$, corresponding to a ZAMS star earlier than O5.5 \citep{han97}, 
independent of its PMS evolutionary stage. 

\subsection{Cluster Population}
In order to examine the assumption that a cluster of embedded young stars is present in the core 
of the molecular cloud, we analyzed the stellar population in two delimited regions; one that we labeled "nebulae" 
that contains the cluster and another that we labeled "control", as ilustrated in Figure 7. 
In that figure we represented the position of all objects detected in the $H$ 
band by crosses. We can see that the small region labeled "cluster region" show a concentration of sources. 
As mentioned in section 3, all the sources with excess emission at 2.2 $\mu$m belong to the nebular region,
which has an area about 1 square arcmin ($\sim 1 \times 1$  sq parsec at 
a distance of 3.8 kpc), compatible with the sizes of other stellar clusters studied  with near infrared arrays 
\citep{persi94,tapia96,persi97}.

In order to separate the cluster sources not detected at $\it{J}$ band from the field stars, we constructed a 
comparative $\it{H-K}_{S}$ 
color distribution diagram for the control and "cluster" regions 
(with the counts for the control region normalized to the "cluster" region area), shown in figure 8 .
From this diagram we can also see the existence of an excess of sources with $H-K_S > 1.0$ in the direction of the nebulae.

Using these results, the following criteria was used to select the "cluster member candidates" for the nebulae 
region: a) sources with $H-K_{S} > 1.0$  that
lie to the right  or on the reddening vector for early type stars in the 
color-color diagram; b) sources with $H-K_{S} > 1.0$ not detected at the $J$ band and 
c) sources detected only at $K_{S}$ band. 
For the b) and c) criteria, the contamination from background stars, mainly late-type giants, should be very small. 
In fact for an average extinction $A_{V}$ of 28 magnitudes, the extinction in the $K$ band $A_K$ is about 3.1
magnitudes (using the interstellar reddening law taken from Rieke \& Lebofsky, 1985), 
and for a distance to the cloud of 3.8 kpc, these sources
would have $K> 16$. 
Table 1 shows the coordinates and photometry  of all selected sources ($J$ and $H$ from both LNA
and 2MASS and $K_S$ and coordinates from 2MASS). 
From the 41 stars with measured $K_S$ magnitudes we were able to measure the LNA magnitudes  in
35 of them at $H$ and 22 at $J$ bands, while the 2MASS survey only detected 20 objects at $H$ and 6 at $J$ bands.

Further information about the nature of the selected objects in Table 1 can be extracted 
from $K_{S}$ versus $(H-K_{S})$ color-magnitude diagram shown in Figure 9.
The locus of the main-sequence 
for class V stars it is also  plotted, with the position of each spectral type earlier than A0 V indicated by 
filled circles. The intrinsic colors were taken from Koornneef (1983) while the absolute $K$ 
magnitudes were calculated from the absolute visual luminosity for ZAMS taken from Hanson et al. 
(1997), for a distance of 3.8 kpc (Caswell \& Haynes). 
The reddening vector for a ZAMS B0 V star, taken from Rieke $\&$ Lebofsky (1985), is shown by 
the dashed line with the positions of  visual extinctions  $A_V = 10$, 20 and 30 magnitudes  
indicated by filled circles. From this diagram we can see that the majority of sources can be 
classified as reddened massive stars earlier than spectral type B5, forming an O-B cluster 
embedded in the molecular cloud. 
However, the spectral types inferred from the color-magnitude diagram are only
upper limits when the stars present emission "excess" in the NIR. 

We made a crude estimate of the star formation efficiency using the mass of gas measured by Karnik et al. (2001).
They mapped the RCW 106 region  at 150  and 210 $\mu$m and detected a compact 
region in the direction of the IRAS  source 16177-5018. 
They modeled the  observed mid and far-infrared SED  assuming a constant gas to dust relation 
$M_{\rm gas}/M_{\rm dust}=100$ and from the integrated luminosity
about $10^5 L_\odot$ they obtained a mass of dust $M_{\rm dust} \approx  30-35 M_\odot$ 
resulting in a mass of gas $M_{\rm gas}\approx  3000-3500\; M_\odot$. 
From this value we  estimated a lower limit to the star formation efficiency ($SFE$), 
using the lower limit to the number of cluster members detected as 41 and an average stellar 
mass of $15 M_\odot$, which corresponds to a  B0 V star \citep{han97}:
\begin{equation}
{\rm SFE} = \frac{M_{\rm stars}}{M_{\rm gas} + M_{\rm stars}} = 0.17
\end{equation}
This is a lower limit because the mass of gas assumed for the nebular region may be 
overestimated since it can include a larger volume than our star forming region and also because
stars fainter than our detection limit are not included.
 
We also estimated the number of Lyman-continuum photons produced in the star forming region,
from the spectral types of the stars that do not show "excess" in the color-color diagram,
using the relation between spectral type and number of Lyman-continuum photons given by Hanson et al. (1997).
We obtained $1.6 \times 10^{50}$ photons s$^{-1}$.

\subsection{The infrared nebula}

In figure 10 we shown contour maps of the infrared nebulae region at $\it{J, H}$ and nb$\it{K}$ bands obtained from our
infrared images. The contours were calibrated in flux with the values starting at 0.18, 0.48 and 
2.8$\times 10^{-4}$ Jy/beam respectively. From the nb$\it{K}$ contour map  
we estimated a lower limit to the flux density of the emission nebula in the $K$ band by measuring the area
between contours and multiplying by the value of the corresponding  
flux density per unit area and we found $S(K)=0.62$ Jy.
Assuming that the emission process is thermal bremsstrahlung from an optically thin
region, we can compare the IR with the radio flux density at 5 GHz given by
Caswell \&  Hynes (1987) to obtain an upper limit to the $K$-band absorption.
Assuming constant density and temperature across the cloud and  local thermodynamic
equilibrium, the flux density $S(\nu)$ due to free-free emission can be written as: 
\begin{equation}
S(\nu)=\tau_\nu B_\nu(T)\Omega
\end{equation}
where $\tau_\nu$ is the optical depth at frequency $\nu$,  $B_\nu(T)$ is the Planck function
\begin{equation}
B_\nu(T)=\frac{2h\nu^3}{c^2}\frac{1}{{\rm exp}(k\nu/kT)-1}
\end{equation}
and 
$\Omega$ is the solid angle of the source given by:
\begin{equation}
\Omega=\pi{(L/2D)}^2
\end{equation}
where $L$ is the diameter of the ionized cloud and $D$ the distance to the observer.
The optical depth at a given frequency $\nu$ is:
\begin{equation}
\tau_\nu=\alpha_{\rm ff}L
\end{equation}
where $\alpha_{\rm ff}$ is the free-free absorption coefficient $\rm (cgs)$ taken from Rybick (1979):
\begin{equation}
\alpha_{\rm ff}=\frac{3.7\times 10^8\;[1-{\rm exp}(-h\nu/kT)]\;(n_e n_i)\;g_{\rm ff}(\nu,T)}{\nu^3\;Z^{-2}\; T^{1/2 }}
\end{equation}
$g_{\rm ff}(\nu,T)$ is the Gaunt factor  obtained from Clayton (1968)
 
For two frequencies $\nu_1$ and $\nu_2$ the ratio of the corresponding flux densities $S(\nu_1)$
and $S(\nu_2)$ may be calculated from:
\begin{equation}
\frac{S(\nu_1)}{S(\nu_2)}={exp}[h(\nu_2-\nu_1)/kT]\frac{g_{\rm ff}(\nu_1,T)}{g_{\rm ff}(\nu_2,T)}
\end{equation}

For a flux density in the 5 GHz continuum of 32.4 Jy and an electron temperature $T_e = 6300$ K, 
as given by Caswell \& Haynes (1987) we find an expected flux density at the $K$ band of 11.5 Jy.
The measured value of 0.62 Jy gives an upper limit to the $K$ band absorption of 3.1 magnitudes.
Using the interestellar extinction law given by Rieke \& Lebofsky (1985) we obtained a visual 
extinction $A_V\approx 27.7$ in very good agreement with the value obtained in section 3.0 
from the median $(H-K)$ color excess.
\par
We observe that is possible that the detected infrared flux at $2.2\mu$m has a contribution from light scattered
by dust, but taking into
account that the ionized region detected at radio wavelengths coincides with the near infrared nebulae and the 
high number
of massive stars inferred from the color-magnitude diagram, we believe that the major contribution to the $\it{K}$
band
integrated flux is probably due to free-free emission from the ionized region.

We also obtained the number of ionizing Lyman continuum photons $N_{Ly}$ from the radio
continuum flux density using the expression derived by Rubin (1968):
\begin{equation}
N_{Ly}= \frac{5.99\times 10^{47}S(\nu) D^2 T_e^{-0.45} \nu^{0.1}}{1+f_i[<He^+/(H^++He^+)]} 
\end{equation}
where  $\nu$ is in MHz and $f_i$ is the fraction of  helium recombination photons that 
are energetic enough to ionize hydrogen ($f_i \approx 0.65$). We  assumed, as in Simpson \& Rubin (1990)
the ratio ${\rm He/H} = 0.1$ 
and  the average volume of $\rm He^+$ as half the volume of ${\rm H^+}$.
Using Equation 7 we found $N_{Ly} \sim 5.2 \times 10^{49}$ photons s$^{-1}$. This value is compatible with the 
number of Lyman continuum photons available from the main sequence stars derived in the previous section.

\subsection{Conclusions}

Near-IR imaging photometry in the direction of the IRAS source 16177-5018 revealed a young massive
stellar cluster and its associated HII region embedded in a dense molecular cloud . We detected 41 cluster member 
candidates concentrated in an area of 1 square parsec, but probably other less massive sources were missed 
because the completeness limit.  

The NIR counterpart of the IRAS point source was identified using more accurate positions from the MSX catalogue. This
source has a near-mid spectral index $d \;{\rm log}(\lambda F_\lambda)/d\; {\rm log} \lambda$  = 4.78 indicating that
the IRAS source is very young, with a lower limit to the bolometric luminosity
inferred from the integrated
near to far-infrared flux density of about $2.8\times 10^5 L_\odot$, which
corresponds to a $ 42\; M_\odot$ star.

The average absorption found for the embedded early type stars ($A_V \sim 28$) coincides with the absorption of the HII
region, derived by comparing the integrated extended flux density at the $K$ band to what is expected
at this band from the observed radio emission.  

The number of
Lyman continuum photons expected from the 11 stars that lie along the reddening line for early-type stars
 about 1.3 $ \times$ 
10$^{50}$ s$^{-1}$, enough to produce the 5 GHz flux density and near IR emission, assumed to be originated by
free-free emission.

\acknowledgments

This work was partially supported by the Brazillian agencies FAPESP and CNPq.
We acknowledge the staff of Laborat\'orio Nacional de Astrof\'isica for their efficient support and to F. Jablonsky 
who was the PI of the CamIV project.
We also thank the anonymous referee for constructive suggestions that improved the quality of this paper.
This publication makes use of data products from the Two Micron All Sky Survey, which is a joint project of the University of 
Massachusets and the Infrared Processing and Analysis Center/California Institute of Technology, funded by the
National Aeronautics and Space Administration and the National Science Foundation.

\clearpage

\begin{deluxetable}{crrrrrrrrrrrrrrrrr}
\tabletypesize{\scriptsize}
\tablecaption{List of the selected near-infrared sources \label{tbl-1}}
\tablewidth{0pt}
\tablehead{
\colhead{IRS}&\colhead{$\alpha$(J2000)}&\colhead{$\delta$(J2000)}&
\colhead{$J_{camiv}$}&\colhead{$\sigma$}& \colhead{$J_{2mass}$}& \colhead{$\sigma$}&
\colhead{$H_{Camiv}$}&\colhead{$\sigma$}&\colhead{$H_{2mass}$}&\colhead{$\sigma$}&\colhead{$K_{2mass}$}&\colhead{$\sigma$}
}
\startdata
1 &16:21:31.60 &$-$50:25:08.3 &16.65 &0.15 &---&--- &12.73 &0.08 &12.76 &0.08 &10.21 &0.08\\
2 &16:21:31.86 &$-$50:25:04.7 &16.49 &0.11 &--- &--- &13.23 &0.09 &13.13 &0.11 &10.92 &0.10\\
3 &16:21:31.35 &$-$50:24:57.9 &--- &--- &--- &--- &14.98 &0.13 &--- &--- &11.25 &0.12\\
4 &16:21:33.03 &$-$50:25:04.6 &--- &--- &--- &--- &15.02 &0.10 &--- &--- &11.37 &0.07\\
5 &16:21:32.21 &$-$50:25:09.6 &--- &--- &--- &--- &15.66 &0.12 &--- &--- &11.39 &0.11\\
6 &16:21:30.77 &$-$50:24:57.1 &--- &--- &--- &--- &15.11 &0.11 &--- &--- &11.44 &0.11\\
7 &16:21:31.34 &$-$50:25:04.1 &17.37 &0.25 &--- &--- &13.88 &0.05 &13.73 &0.14 &11.45 &0.09\\
8 &16:21:30.56 &$-$50:25:19.2 &18.07 &0.30 &--- &--- &14.23 &0.09 &14.14 &0.13 &11.56 &0.12\\
9 &16:21:33.94 &$-$50:24:31.5 &16.39 &0.11 &--- &--- &13.27 &0.05 &13.31 &0.04 &11.66 &0.04\\
10 &16:21:31.12 &$-$50:24:51.9 &17.30 &0.27 &--- &--- &14.87 &0.08 &--- &--- &11.68 &0.13\\
11 &16:21:30.45 &$-$50:25:04.2 &15.75 &0.13 &15.65 &0.10 &13.46 &0.05 &13.33 &0.11 &11.73 &0.13\\
12 &16:21:30.52 &$-$50:25:14.5 &16.15 &0.14 &16.19 &0.11 &14.70 &0.06 &14.68 &0.12 &11.89 &0.12\\
13 &16:21:30.89 &$-$50:25:07.2 &17.56 &0.32 &--- &--- &14.50 &0.06 &14.53 &0.13 &11.96 &0.09\\
14 &16:21:31.01 &$-$50:25:26.0 &--- &--- &--- &--- &15.54 &0.12 &--- &--- &12.09 &0.13\\
15 &16:21:31.10 &$-$50:25:15.1 &--- &--- &--- &--- &14.72 &0.08 &--- &--- &12.24 &0.15\\
16 &16:21:28.02 &$-$50:24:42.6 &15.21 &0.08 &15.23 &0.06 &13.30 &0.05 &13.34 &0.05 &12.39 &0.05\\
17 &16:21:30.38 &$-$50:25:25.1 &--- &--- &--- &--- &15.86 &0.10 &--- &--- &12.46 &0.17\\
18 &16:21:28.79 &$-$50:24:09.9 &16.43 &0.14 &--- &--- &13.97 &0.07 &13.92 &0.06 &12.50 &0.07\\
19 &16:21:31.84 &$-$50:24:45.9 &--- &--- &--- &--- &15.35 &0.07 &15.32 &0.18 &12.51 &0.10\\
20 &16:21:29.53 &$-$50:25:00.9 &16.69 &0.13 &--- &--- &14.59 &0.08 &--- &--- &12.51 &0.10\\
21 &16:21:30.09 &$-$50:25:23.3 &16.72 &0.11 &--- &--- &15.28 &0.06 &--- &--- &12.64 &0.14\\
22 &16:21:29.50 &$-$50:24:44.9 &15.29 &0.08 &15.21 &0.06 &13.57 &0.03 &13.66 &0.03 &12.73 &0.08\\
23 &16:21:29.39 &$-$50:24:49.4 &16.94 &0.17 &--- &--- &14.24 &0.07 &14.44 &0.06 &12.82 &0.03\\
24 &16:21:30.91 &$-$50:24:40.7 &17.71 &0.28 &--- &---  &15.07 &0.04 &15.08 &0.13 &12.88 &0.16\\
25 &16:21:26.42 &$-$50:24:47.0 &16.87 &0.13 &--- &--- &14.38 &0.05 &14.25 &0.11 &13.09 &0.09\\
26 &16:21:28.48 &$-$50:24:34.3 &16.45 &0.08 &--- &--- &14.39 &0.05 &14.34 &0.10 &13.10 &0.11\\
27 &16:21:29.46 &$-$50:25:16.7 &15.59 &0.08 &15.76 &0.07 &14.60 &0.05 &14.53 &0.08 &13.45 &0.18\\
28 &16:21:29.64 &$-$50:24:02.4 &16.63 &0.13 &--- &--- &14.64 &0.05 &14.74 &0.15 &13.57 &0.11\\
29 &16:21:31.68 &$-$50:25:31.8 &--- &--- &--- &--- &15.45 &0.07 &--- &--- &13.58 &0.14\\
30 &16:21:33.41 &$-$50:25:18.5 &--- &--- &--- &--- &15.56 &0.06 &--- &--- &13.63 &0.06\\
31 &16:21:28.92 &$-$50:25:00.1 &16.71 &0.11 &--- &--- &14.73 &0.05 &14.80 &0.13 &13.73 &0.13\\
32 &16:21:34.11 &$-$50:25:27.9 &--- &--- &--- &--- &16.52 &0.11 &--- &--- &13.80 &0.15\\
33 &16:21:27.95 &$-$50:26:18.5 &--- &--- &--- &--- &14.97 &0.07 &--- &--- &13.93 &0.11\\
34 &16:21:33.66 &$-$50:25:31.3 &16.01 &0.13 &15.95 &0.10 &15.25 &0.10 &15.10 &0.14 &14.04 &0.18\\
35 &16:21:34.86 &$-$50:24:36.6 &--- &--- &--- &--- &17.35 &0.20 &--- &--- &14.26 &0.09\\
36 &16:21:27.82 &$-$50:26:21.7 &--- &--- &--- &--- &--- &---  &--- &--- &13.38 &0.13\\
37 &16:21:33.51 &$-$50:25:46.5 &--- &--- &--- &--- &--- &---  &--- &--- &13.51 &0.16\\
38 &16:21:33.92 &$-$50:25:41.4 &--- &--- &--- &--- &--- &---  &--- &--- &13.56 &0.12\\
39 &16:21:35.38 &$-$50:24:59.6 &--- &--- &--- &--- &--- &---  &--- &--- &13.89 &0.05\\
40 &16:21:35.95 &$-$50:26:08.4 &--- &--- &--- &--- &--- &---  &--- &--- &14.11 &0.07\\
41 &16:21:28.43 &$-$50:25:35.6 &--- &--- &--- &--- &--- &---  &--- &--- &14.87 &0.15\\
 \enddata


\end{deluxetable}

\begin{figure*}
\plotone{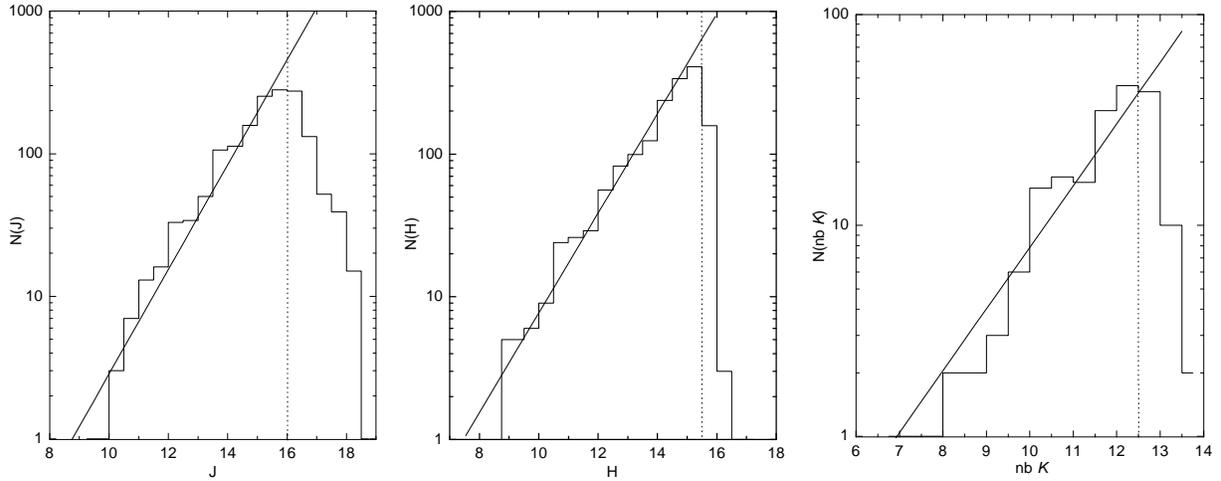}
\caption{Histograms of $J$ (a), $H$ (b) and $nbK$ (c) magnitude counts. The completeness limits are indicated by the
vertical dotted lines.\label{fig1}}
\end{figure*} 

\begin{figure*}
\plotone{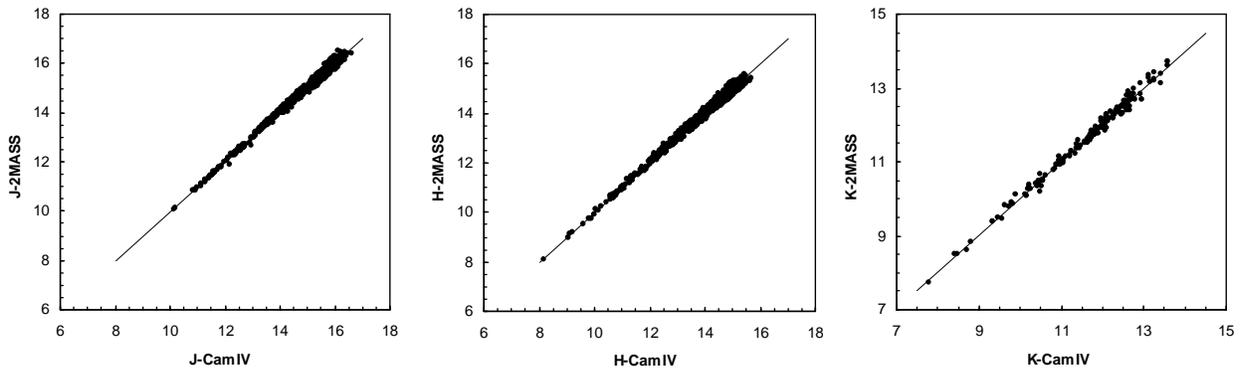}
\caption{The magnitudes comparative diagram M$_{JHK}$(2MASS) $\times$ M$_{JHK}(CamIV)$. The continuous line shows the relation expected
if the two photometric systems were equal.  \label{fig2}}
\end{figure*}


\begin{figure*}
{\includegraphics{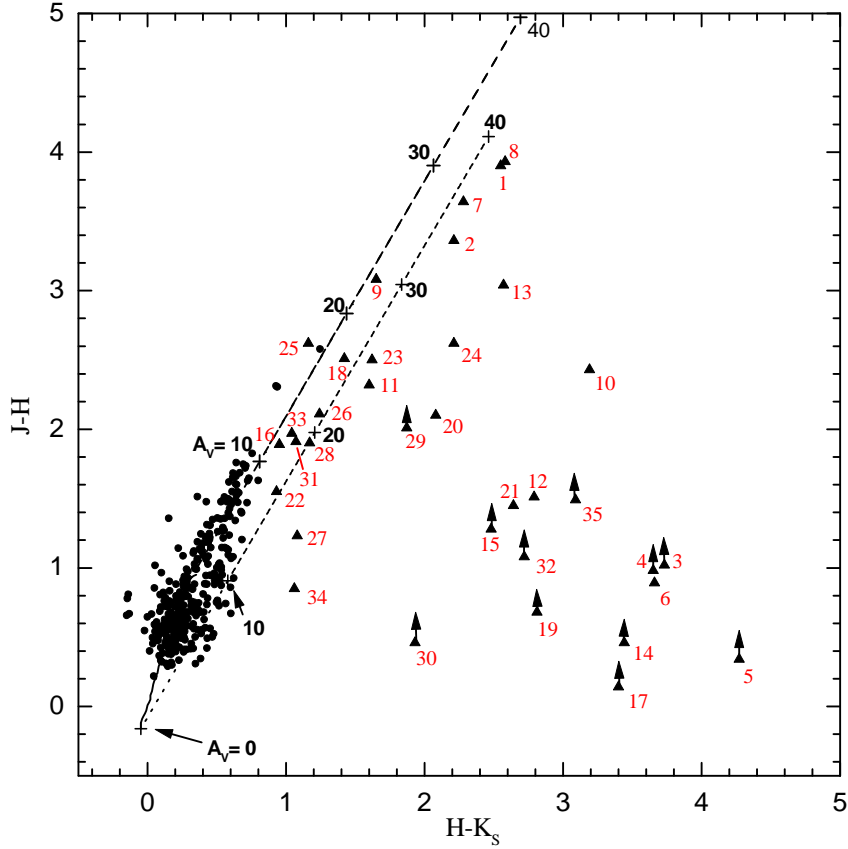}}
\caption{Color-color diagram of all sources detected at the three passbands. The locus of the main sequence
and giants branch are shown by the continuous lines taken from Koornneef (1983),
while the two parallel lines (dashed and dotted) follow the reddening vectors taken from Rieke $\&$ Lebofsky (1985).
The location (crosses) of $A_V$ = 0, 10, 20, 30 and 40 magnitudes of visual extinction are indicated by black bold numbers. We also plotted the 
sources only detected at H and 2MASS K$_{S}$ images using the completenees limit of the J photometry (about 16 mag).
These sources have lower limit to the J-H color  indicate by arrows. The cluster members candidates taken from table
1 are labeled by the red numbers.\label{fig4}}
\end{figure*}

\begin{figure*}
{\epsscale{0.9}\plotone{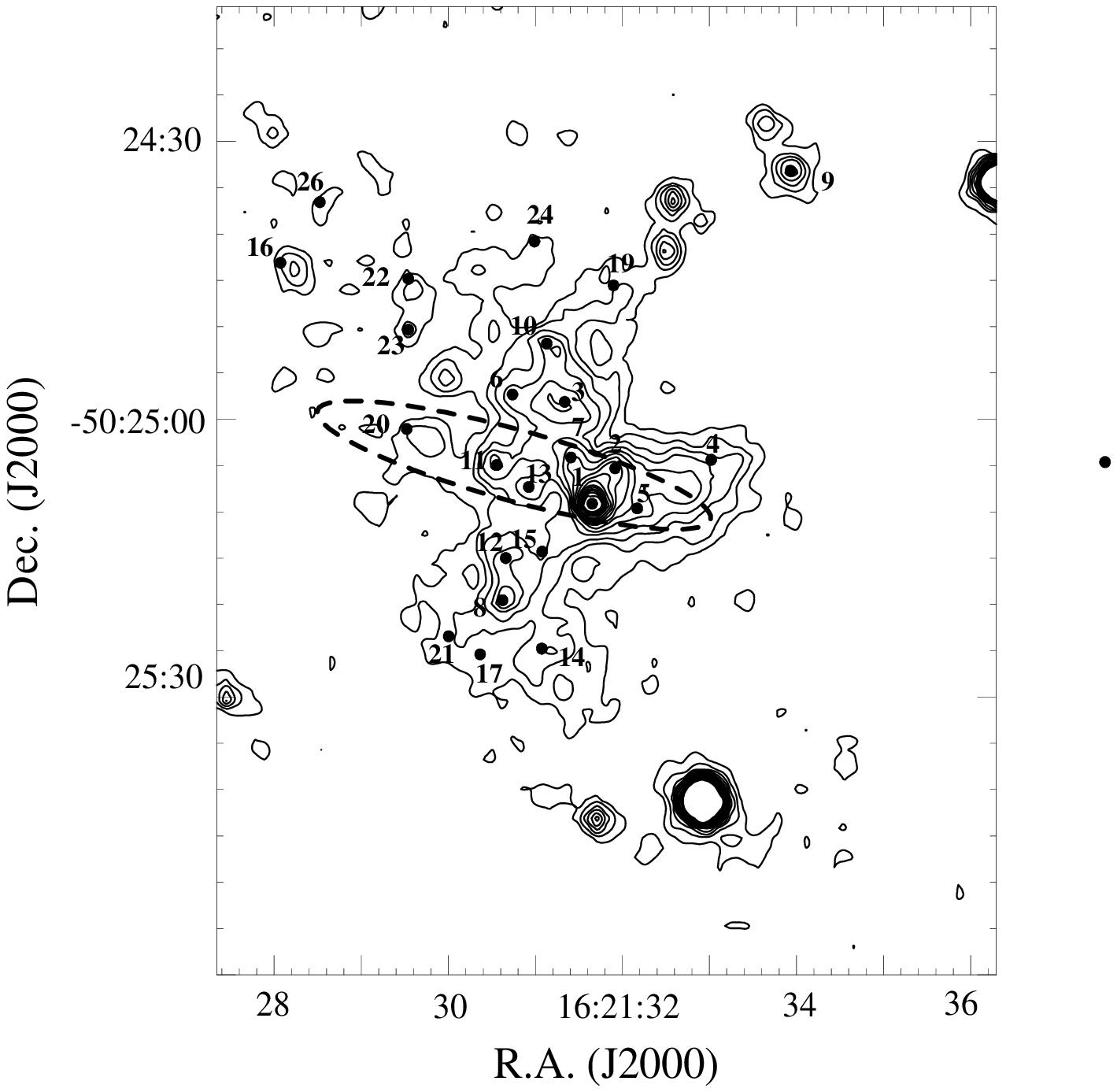}}
\caption{nbK band contour map of the infrared nebulae associated with IRAS16177-5018.
 The contours start at 1.8$\times 10^{-4}$ Jy/beam, with the same intervals (the beam size is $2 \times 2$ \ pixels).
The positions of selected infrared sources taken from table 1 are indicated by black numbers.
It is also indicated the location of the IRAS coordinate elipse error (black dotted line).\label{fig5}}
\end{figure*}

\begin{figure*}
{\includegraphics{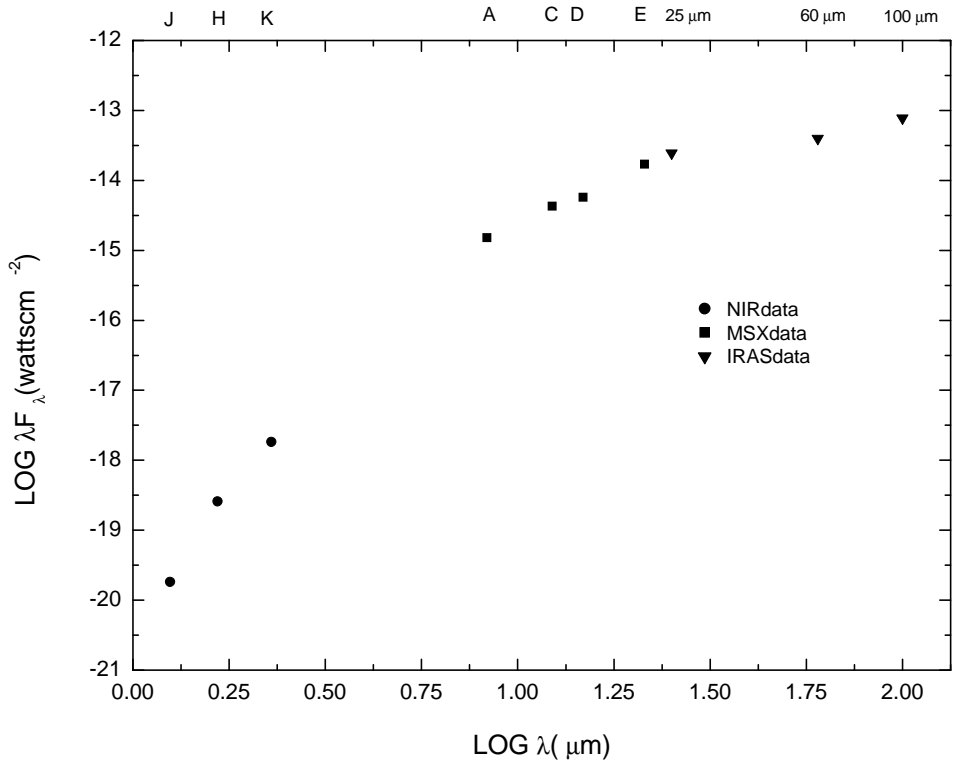}}
\caption{Spectral energy distribution of the IRS7 infrared source. The flux for the three near infrared bands (filled 
circles) were taken from our survey. The mid infrared data (filled squares) were taken from $MSX$ $catalogue$
(bands A=8.28$\mu$m, C=12.13$\mu$m, D=14.65$\mu$m and E=21.34$\mu$m) while the far infrared data 
(filled down triangles) were taken from IRAS point source catalogue.\label{fig6}}
\end{figure*}

\begin{figure*}
{\includegraphics{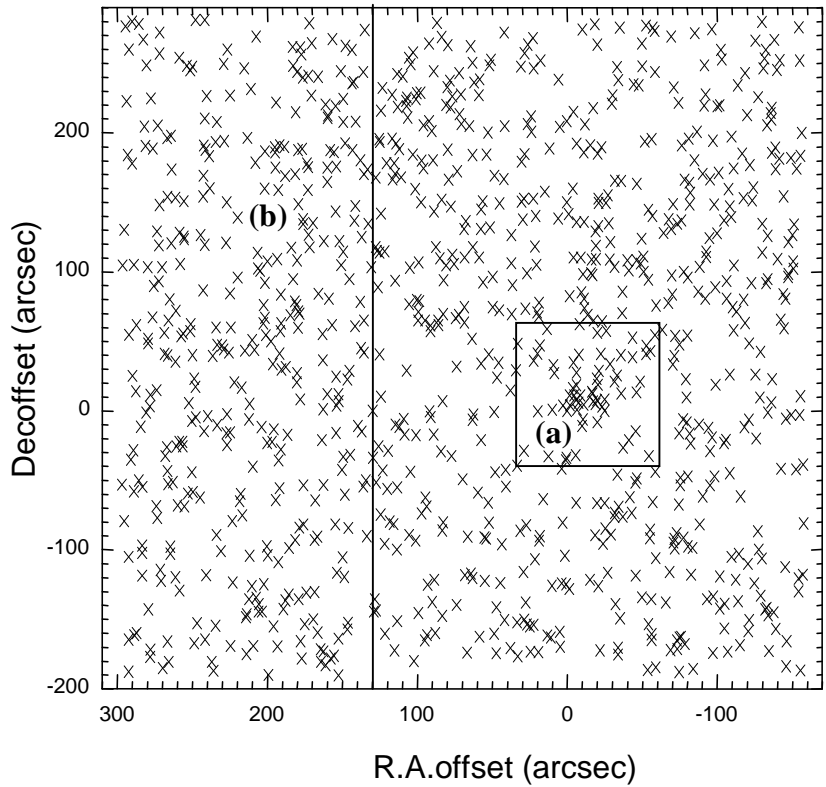}}
\caption{Diagram of the spacial distribution of sources detected at LNA's $\it{H}$ band . The two comparative regions
{\bf a} (nebulae) and {\bf b} (control) are delimitated by the vertical continuum line. The "cluster region" area is delimited
by the dotted square at the up right side.\label{fig7}}
\end{figure*}

\clearpage

\begin{figure*}
{\epsscale{0.7}\plotone{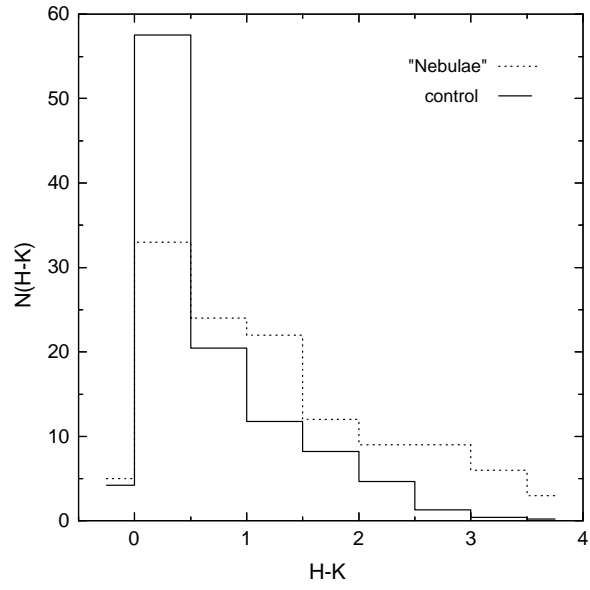}}
\caption{Comparative diagram of the distribution of the $\it{H-K}$ color for the "control" (continuous line) and
"nebulae" (dotted line) areas  .\label{fig9}}
\end{figure*}

\begin{figure*}
{\epsscale{0.95}\plotone{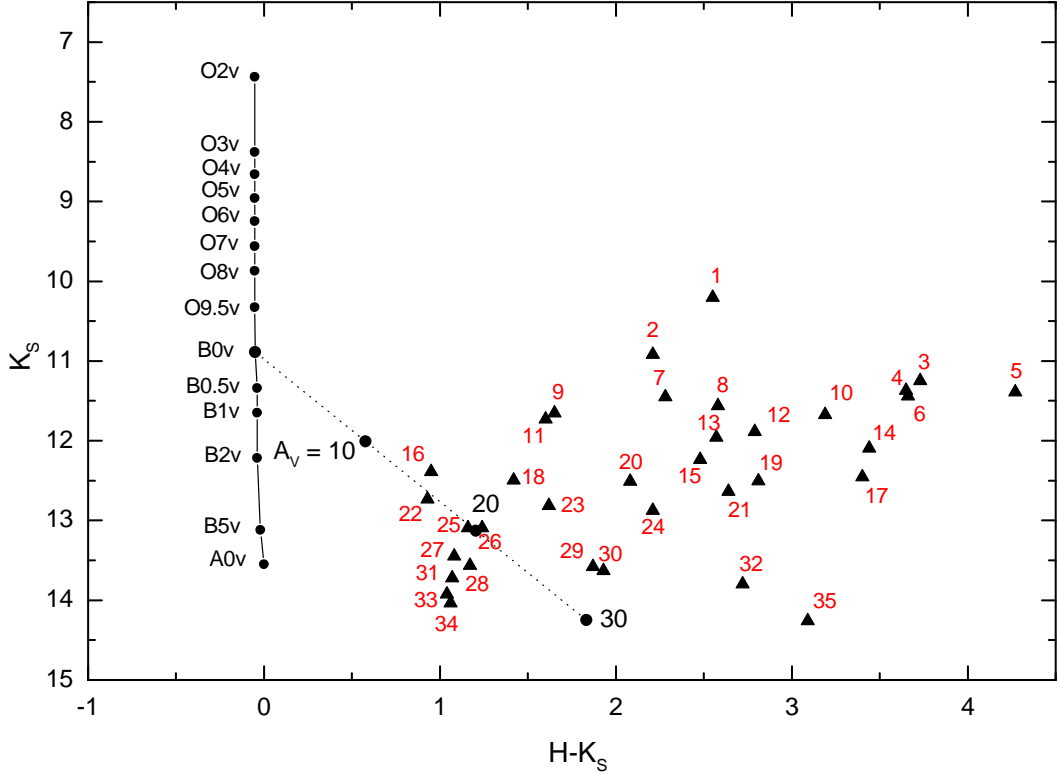}}
\caption{$K_{S} \times (H-K_{S})$ color-magnitude diagram of the selected sources in table 1.
The locus of the main sequence (dwarfs) at 3.8 kpc is shown by the continuous line. The intrinsic colors were taken
from Koornneef (1983) while the absolute $\it{K}$ magnitudes were calculated from the absolute visual luminosity for ZAMS taken from
Hanson et al. (1997). The reddening vector for a B0 ZAMS star (dotted line) was taken
from Rieke $\&$ Lebofsky (1985); Also are indicated the location (filled circles) of $A_V$ = 10, 20 and 30 
magnitudes of visual extinction. The sources IRS1-IRS35 are labelled by red numbers.\label{fig10}}
\end{figure*}

\clearpage

\begin{figure*}
{\epsscale{1}\plotone{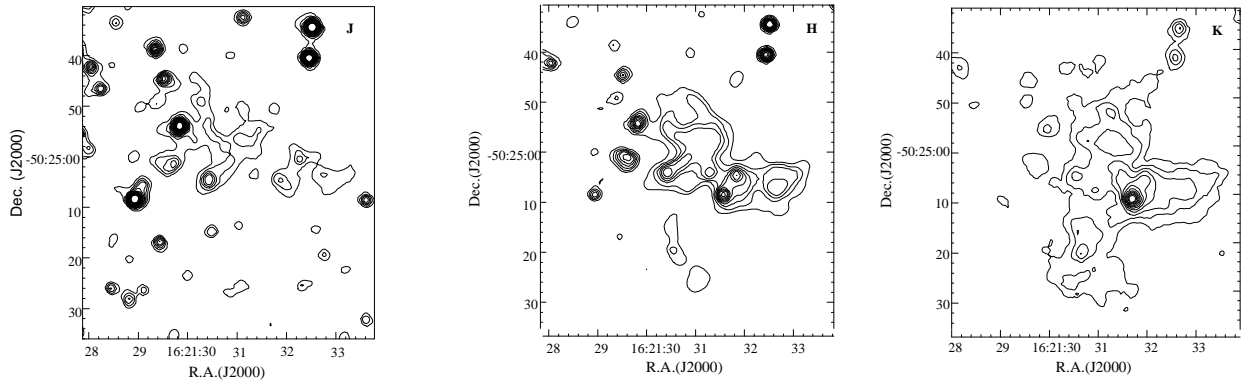}}
\caption{ J, H and nbK contour maps of the infrare nebulae region. The contours start at 1.8$\times 10^{-5}$ Jy/beam in $\it{J}$,
4.8$\times 10^{-5}$ Jy/beam in $\it{H}$ and 2.8$\times 10^{-4}$ Jy/beam in nb$\it{K}$, and are spaced by the 
same fluxes per beam (a beam is 2$\times$2 pixels).\label{fig11}}
\end{figure*}

\end{document}